\documentclass[14pt,onecolumn]{emulateapj}
\usepackage{epsf}

\def\ltsima{$\; \buildrel < \over \sim \;$}
\def\lsim{\lower.5ex\hbox{\ltsima}}

\def\deg{\ifmmode{^\circ}\else{$^\circ$}\fi}
\def\hGpc{\ifmmode{h^{-1}{\rm Gpc}}\else{$h^{-1}{\rm Gpc}$}\fi}
\def\hkpc{\ifmmode{h^{-1}{\rm kpc}}\else{$h^{-1}{\rm kpc}$}\fi}
\def\hMpc{\ifmmode{h^{-1}{\rm Mpc}}\else{$h^{-1}{\rm Mpc}$}\fi}
\def\hMsun{\ifmmode{h^{-1}M_\odot}\else{$h^{-1}M_\odot$}\fi}

\def\muK{\ifmmode{\mu{\rm{K}}}\else{$\mu$K}\fi}
\def\mum{\ifmmode{\mu{\rm{m}}}\else{$\mu$m}\fi}

\usepackage{ulem}
\usepackage{color}

\begin{document}

\title{Short GRBs and dark matter seeding in neutron stars}

\author{M. \'Angeles P\'erez-Garc\'ia$^1$~\footnote{mperezga@usal.es}, F. Daigne$^2$~\footnote{daigne@iap.fr} and  J.  Silk$^{2,3}$~\footnote{j.silk1@physics.ox.ac.uk} }

\affil{$^1$ Departament of Fundamental Physics and IUFFyM,\\ University of  Salamanca, 
Plaza de la Merced s/n 37008 Salamanca, Spain\\
$^2$ UPMC-CNRS, UMR7095, Institut d'Astrophysique de Paris, 75014 Paris, France\\
$^3$ Oxford Physics, University of Oxford, Keble Road OX1 3RH, Oxford, United Kingdom}

\begin{abstract}
We present a mechanism 
based on internal self-annihilation of dark matter accreted from the galactic halo in the inner regions of neutron stars that may trigger the full or partial conversion into a quark star. We explain how this effect may induce a gamma ray burst that could be classified as short, according to the usual definition based on time duration of the prompt gamma-ray emission. 
This mechanism differs in many aspects from the most discussed scenario associating short gamma-ray bursts  with compact object binary mergers. We list possible observational signatures that should help to distinguish between these two possible classes of progenitors.
\end{abstract}



\section{Introduction}
Gamma ray bursts (GRBs) are interesting highly energetic phenomena for which  there is still no definitive explanation on the originating mechanism \citep[see e.g.][]{piran:04,gehrels:09}. 
Regarding the initial event that triggers a GRB, the combination of the observed short variability timescales and huge radiated energies points towards cataclysmic events leading to the formation of a stellar compact object with mass $M$ and radius $R$  releasing a gravitational energy  $\Delta E \approx G M^2 / {R}\approx 10^{53}-10^{55}$ erg.
Specifically, according to their time duration, GRBs can be classified as ${\it long}$ (LGRBs) if 
the duration of the gamma-ray signal is larger than 2 s
and short (SGRBs) in other cases \citep{kouveliotou:93}, although the boundary is not sharply delimited 
\citep[see e.g.][]{zhang:09,bromberg:12}.
There are several pieces of evidence in favour of an association of LGRBs with massive stars, especially the facts that they occur in star-forming regions \citep{bloom:2002} and that supernovae have been found in association with a few of them \citep[see e.g.][]{grb-SN}. This leads to the {\it collapsar} model \citep{woosley:93} where a LGRB is associated with the gravitational collapse of a massive, Wolf-Rayet, star leading to the formation of  an accreting stellar-mass black hole (BH). 
An alternative scenario where the central engine formed after the gravitational collapse is a young, rapidly rotating (ms period) neutron star (NS) has also been considered by \cite{usov:92}, \cite{thompson:94} and \cite{metzger:11}.
On the other hand, the identification of the progenitors of SGRBs is more uncertain \citep{nakar:07}. 
This is partially due to the fact that the detection of SGRB afterglows is technically more difficult, and therefore rarer, than for LGRBs. So far there is only a handful of detected afterglows, since the first detection in 2005 \citep{SGRB-afterglow,villasenor:05}. 
SGRBs can occur in all types of galaxies, either of a star-forming type or not \citep{berger:11}. This is consistent with the most popular scenario, that is 
the merger of a NS+NS or NS+BH binary system.

As a main motivation for the present contribution and due to the current uncertainties on the origin of (especially short) GRBs, it is worth  investigating other possible classes of cataclysmic events where astrophysical stellar compact objects are involved. We will focus on NSs. 
A NS can be partially characterized by a set of static observables, namely mass,  $M_\mathrm{NS}$, and radius, $R_\mathrm{NS}$. The structure of a NS has been discussed thoroughly in the past \citep[for a review see e.g.][]{ns-structure}.
It can be briefly described 
by a dense {\it core}, where matter is in the form of a liquid of nucleons (or additional particles including strangeness), and a {\it crust} where matter undergoes a liquid-solid transition into a clusterized myriad of nuclei \citep{pasta}. In  more detail, the so-called neutron drip (ND) line signals the boundary between the outer and inner crust where neutrons start to leak out from nuclei. The mass in the crust is small (up to a few percent) as compared to the whole NS and therefore it could be accelerated, if ejected by large energies, to large velocities or even high Lorentz factors.

The equation of state (EoS) that governs the description of matter in the inside of NSs 
is currently poorly known. Whether matter remains in the form of  baryons or deconfined quarks is a matter of debate. The problem of the internal constituents in a NS has been thoroughly and periodically revisited from the early 1930's when Landau predicted the existence of such objects \citep{landau} from the basis of the existence of neutral hadronic particles. From the discovery  of the neutron by Chadwick in 1932  to 1984 when it was hypothesized \citep{witten} that a more stable type of matter could be possible (formed by roughly equal quantities of $uds$ quarks, arising from weak decay of regular $ud$ matter present in nucleonic matter made of protons and neutrons) a wide variety of possible nuclear configurations has been proposed. In this latter quark system case, for example, recent attention has been devoted to lattice calculations showing the possibility of the existence of the $H-$dibaryon, with  $uuddss$ quark content \citep{hdibaryon} and  experimental input from the LHC is expected on the formation of a quark-gluon plasma.  

A stable type of extended strange matter would allow the existence of stars even more compact than NSs, i.e. hybrid or pure quark stars (QSs) \citep{alcock:86}. The transition from a nucleonic type of matter to a quark phase is  presently not well understood. From the theoretical point of view, some work has been done \citep{nucleation1, nucleation2} on the possibility of quark bubble nucleation in a cold system, where the temperature is well approximated by $T\approx 0$. As this mechanism is not efficient enough, this process may seem unlikely to occur. Another series of works have been developed by Ouyed and collaborators which propose the Quark-Nova model in which a massive NS converts explosively to a quark star \citep{ouyed02}. The transition is obtained mostly via a central density increase due to spin-down or accretion forming a conversion front that propagates
toward the surface. The outcome is ejection of the NS’s outermost layers at relativistic speeds. Alternatively, as we explain in this work, additional {\it external} agents may be able to produce thermal excitations in these dense degenerate matter systems.

 From indications concerning astrophysical and cosmological backgrounds we know that dark matter (DM) is distributed inhomogeneously in the universe and, additionally, could be gravitationally accreted onto individual massive astrophysical objects.  One of the most popular dark matter candidates is considered to be a  Majorana-type weakly interacting particle (WIMP), although other asymmetric dark matter candidates could exist as well \citep{asym}. From the experimental point of view, current direct and indirect detection experiments try to constrain its properties. For example, direct detection earth-based experiments DAMA \citep{Dama}, CoGeNT \citep{Cogent}, CRESST  and lately COUPP \citep{Behnke:12} seem to point towards a light particle in the $\sim 10-20$ $\rm GeV/c^2$ mass range (although this is in tension with other null searches) while there has been a lot of interest in indirect detection of new excesses of extra photon components from the Galactic Center that could be due to annihilation of a $\sim 130$ $\rm GeV/c^2$  particle \citep{fermi1, fermi2}, albeit so far at only the $2-3\sigma$ level of significance.

 DM passing through astrophysical massive objects is  expected to scatter off their constituent nuclei, lose energy, and become trapped by their gravitational field. Multiple scatterings cause DM to sink towards the center and afterwards, assuming that it may self-annihilate, a series of end products are available. Based on the previous arguments, a set of current experiments try to find evidence for this process in the sun, earth \citep{baikal,baer} or even more compact-sized objects such as  NSs \citep{kou}. 

Recently, new ideas on the possibility that quark bubbles may be formed by spark seeding in the inner cores of dense NSs have been proposed by \cite{perez:10}. Sparks would be generated in a sort of {\it Trojan horse} mechanism by self-annihilation of dark matter particles gravitationally accreted by NSs gathering in their internal cores. As estimated, this would allow release energies of up to about one order of magnitude higher than the quark binding energies in the present baryons \citep{perez:12}. This mechanism then could lead to the conversion of NSs into QSs if the transition to quark matter phase is macroscopic. In this way the matter component in the form of DM would not only influence cosmological observables, but even NS dynamics \citep{perez:11}. Similar arguments have been claimed in the solar seismology observables \citep{lopes}.

We investigate in the present paper whether this type of {\it non-repeating} cataclysmic event could produce SGRBs and what could be a possible specific signature. After a short summary of the scenario in \S\ref{sect:scenario}, in \S\ref{sect:energetics} we compare the energy released by SGRBs with the energetics of the progenitor model that we propose and in \S\ref{sect:rate} we discuss the short GRB event rate in light of this scenario as compared to 
observed rates, as well as the natural delay time between the regular NS phase and  QS formation and how this mechanism may affect the properties of the host galaxies of short GRBs. In  \S\ref{sect:engine} we present and analyze the ejecta and crust masses that could be expelled due to the NS$\to$QS conversion and we present the details of the resulting central engine model.
In \S\ref{sect:grb} we attempt to discuss the expected temporal (\S\ref{sect:duration}) and spectral (\S\ref{sect:spectrum}) properties of the prompt gamma-ray emission and the afterglow.
In \S\ref{sect:gw} we discuss our results and possible genuine additional signals and gravitational wave emission that could finally lead to a clear identification as compared with the short GRB scenario 
involving binary mergers. Finally in \S\ref{sect:conclusions} we give a brief summary of our conclusions and suggestions on how to proceed in further work.


\section{Summary of the scenario} 
\label{sect:scenario}
There has been some previous work where DM accretion in astrophysical objects has been considered \citep{nusi, kou, tin}. As a summary, we start in this section by  motivating the scenario where the emission of 
a SGRB may occur.

 In our previous work \citep{perez:10} the possibility of  spark seeding in NS as a result of DM accretion was proposed.  We found a relationship between the mass of the DM particle, $m_{\chi}$, and the binding energy of the lump of quark matter or strangelet formed, $E^A_{\rm slet}(\mu_i (n_A), m_i, B)$ as calculated from the MIT bag model \citep{mit}. 
\begin{equation}
2 f m_{\chi} c^2 \geq E^A_{\rm slet}(\mu_i (n_A), m_i, B),
\label{ineq}
\end{equation}
where $n_A$ is the baryonic number density and $\mu_i$ and $m_i$ are the chemical potential and mass of the quark of $\rm ith-$type in the light sector, $i=u,d,s$. $B$ is the bag constant. The channels available to the ${\chi}$ annihilation are, so far, not known and this uncertainty is phenomenologically parametrized into the fraction $f$. Inside the NS, it is believed that DM scatters one or more times and thermalizes according to a Boltzmann distribution. The thermal radius is given by 
\begin{equation}
r_\mathrm{th}(t)=\left( \frac{3 \,k_\mathrm{B} T_\mathrm{c}(t)}{2 \pi G \rho_\mathrm{c}(t) m_{\chi}}\right)^{1/2} .
\end{equation}
This radius is, in general, time-dependent since evolution proceeds for central temperatures $T_\mathrm{c}$ and baryonic mass densities $\rho_\mathrm{c}$. DM  at high density may self-annihilate to channels involving photons, leptons and light $q{\bar q}$ pairs \citep{kuhlen}. Due to the fact that either a light or heavy self-annihilating particle in the $\sim 1-100 \,\rm GeV/c^2$ mass-range may release a spark of energy able to liberate the constituent quark content of the baryons forming the very dense matter in the core of these objects, a pure nucleonic NS could then be considered a metastable state of compact stars. Note that a self-annihilating $\sim 20$ $\rm GeV/c^2$ particle as recently suggested in the context of direct detection experiments would, in particular, fit into this scenario. The  annihilation power is then given as 
\begin{equation}
{\cal L}(r)=\frac{<\sigma_\mathrm{a} v> c^2}{m_{\chi}} \int_0^r \rho^2_{\chi}(r') d^3r'\, ,  
\end{equation}
where $\rho_{\chi}$ is the DM mass density and $<\sigma_\mathrm{a} v> \simeq 3\,\times\,10^{-26}\rm cm^3/s$ is the thermally averaged annihilation cross section.

If the spark seeding is able to form quark lumps or strangelets, most likely off-center,  inside a volume $\sim r^3_\mathrm{th},$ they could individually grow or coalesce from a stable minimum mass number, so far poorly known, $A_\mathrm{min}\sim 10-100$, according to previous estimates \citep{madsen-min}. In this way a {\it burning front} may proceed if particular thermodynamical conditions are fulfilled. If the burning is partial, the final configuration object would be hybrid, constitued by an  inner quark matter phase and an outer nucleonic phase. However, if the burning is complete the resultant object would be formed by a pure quark phase. In both cases the final configuration in the mass-radius phase space $(M_\mathrm{F}, R_\mathrm{F})$, as computed from different
 EoS using a variety of many-body effective theories \citep{eos}, is a smaller, slightly lighter, more compact object than the initial one characterized by $(M_\mathrm{i}, R_\mathrm{i})$. 

\begin{figure}
\begin{center}
\includegraphics [angle=-90,scale=0.75] {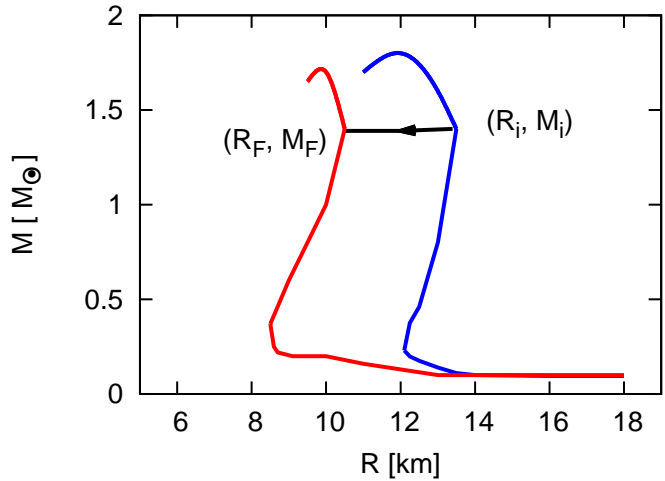}
\caption{Qualitative scheme of transition from an initial configuration $(M_\mathrm{i}, R_\mathrm{i})$ to a final one with $(M_\mathrm{F}, R_\mathrm{F})$. In blue (red) we show different compact star behaviour for NS (hybrid QS) adapted from \citet{evolutionNSQS}. Black line depicts the possible transition from one configuration to another, which could be triggered by internal self-annihilation of DM. 
}
\label{fig:mr}
\end{center}
\end{figure}

In  Figure~\ref{fig:mr} we show a qualitative scheme, inspired from the work by  \citet{evolutionNSQS}, of how the transition would produce a more compact object.
The transition point, and therefore the initial configuration mass, would be influenced by the presence of spark seeding by self-annihilating DM in this model.

In such a NS $\to$ QS transition, part of the outer crust of the initial NS can be expelled, and possibly accelerated to relativistic speed, which may lead to a transient episode of high-energy emission. 
In the following, we discuss a possible association of some GRBs with such events. We focus on SGRBs for several reasons:
\begin{itemize}
\item the energetics of bright long GRBs may be a challenge for this scenario, whereas typical energies of short GRBs are in a much better agreement (see \S\ref{sect:energetics});

\item a delay of at least $\Delta\sim10^3$ yr is expected between the formation of a NS and the transition to a QS, which is too long for long GRBs as supernovae within a few days have been found in association with some of these events.
On the other hand, the scenario naturally leads to a broad distribution of the delay $\Delta t$, allowing long delays that are necessary to explain the properties of the afterglows and host galaxies of short GRBs (see \S\ref{sect:rate});

\item the scenario naturally leads to short (or very short) timescales for the gamma-ray emission, pointing towards short or ultra-short GRBs (see \S\ref{sect:duration}).
\end{itemize}


\section{Energetics} 
\label{sect:energetics}

Since 2005 and the first accurate localization of a SGRB \citep{SGRB-afterglow}, the redshift of an increasing number of short GRBs has been measured, confirming their cosmological origin and leading to a large dispersion in the isotropic equivalent energy released in gamma-rays, $E_\mathrm{\gamma,iso}\simeq 10^{48}-10^{52}$ erg \citep{nakar:07,berger:07}. 
This energy is emitted by a relativistic outflow which is probably beamed within an opening angle $\theta_\mathrm{j}$ so that the true energy release is $E_\mathrm{\gamma}=f_\mathrm{b}^{-1} E_\mathrm{\gamma,iso}$, where the beaming factor is defined as
\begin{equation}
f_\mathrm{b}=\left(\frac{\Omega}{4\pi}\right)^{-1}=\left(1-\cos{\theta_\mathrm{j}}\right)^{-1}\simeq \frac{2}{\theta_\mathrm{j}^2}\,\, \, \, \, \,  \mathrm{if}\, \theta_\mathrm{j}\ll 1\, .
\end{equation}
Unfortunately, the opening angle $\theta_\mathrm{j}$ cannot be easily measured in GRBs. It can be estimated if a \textit{jet break} is identified in the late afterglow \citep{rhoads:97}. It is only in very rare cases that a precise constraint on $\theta_\mathrm{j}$ is obtained, such as $\theta_\mathrm{j}\simeq 7^\circ$ in GRB 051221A \citep{soderberg:06}, and even in such cases, it remains partially model-dependent. The distribution of $f_\mathrm{b}$ for SGRBs is therefore unknown \citep{nakar:07}. Highly beamed SGRBs with $\theta_\mathrm{j}\simeq 1^\circ$ would lead to $f_\mathrm{b}=6.5\times 10^3$ whereas larger opening angles of about $\theta_\mathrm{j}\simeq 30^\circ$ give $f_\mathrm{b}=7.5$. This will remain as a major source of uncertainty in what follows.  

In SGRBs, the isotropic equivalent energy $E_\mathrm{\gamma,iso}$ is released in gamma-rays on a very short time scale ($\la 2$ s) and observed well above 1 MeV \citep[see e.g.][]{guiriec:10}. Then, the gamma-ray emission must be produced from relativistic ejecta to avoid the \textit{compactness problem}: this provides a  constrain on the minimum value of the Lorentz factor for the emission region to be optically thin for $\gamma\gamma$ annihilation \citep{rees:66, baring:97, lithwick:01}. The analysis of a few pre-\textit{Fermi} short bursts \citep{ghirlanda:04} leads to $\Gamma > \Gamma_\mathrm{min}\simeq 15$ from this $\gamma\gamma$ opacity constraint \citep{nakar:07}. When a SGRB is detected up to the GeV range by \textit{Fermi}-LAT, stricter constraints are obtained, such as $\Gamma\ga 300-1000$ in GRB 090510 \citep{ackermann:10,hascoet:12}.\\

Let us compare these constraints on the energetics and the relativistic nature of the emitting material in SGRBs with the predictions of the scenario described in the previous section. If there is a transition triggered from the inside \citep{perez:10}, namely due to DM self-annihilation in the central core of a NS ($M_\mathrm{NS}$, $R_\mathrm{NS}$) then the final configuration may lead to a more compact object with partially deconfined quark content. The energy gap to this new configuration ($M_\mathrm{QS}$, $R_\mathrm{QS}$) will release approximately (here we assume $M_\mathrm{NS} \approx M_\mathrm{QS}$ for the sake of comparison),
\begin{equation}
\Delta E_\mathrm{grav}\simeq G M_\mathrm{NS}^2\left(\frac{1}{R_\mathrm{QS}}-\frac{1}{R_\mathrm{NS}}\right).
\end{equation}
Note that the internal energy due to the matter equation of state has been estimated to be of the order of $ E_\mathrm{grav}$ \citep{bombaci}, therefore this estimate is correct up to a numerical factor that somewhat depends on the EoS. 
The  energy $\Delta E_\mathrm{grav}$ liberated in a NS$\to$QS transition can be compared with the energy $\Delta E_\mathrm{grav,prog\to NS}\simeq G M_\mathrm{NS}^2/R_\mathrm{NS}$ released in the gravitational collapse leading to the formation of a NS (standard core-collapse supernova). We define this ratio (in the approximation $M_\mathrm{NS}\approx M_\mathrm{QS}$) as $f_\mathrm{QS/NS} \approx \left| \Delta E_\mathrm{grav,NS\to QS}/\Delta E_\mathrm{grav,prog \to NS} \right| \approx \left| \left(\frac{R_\mathrm{NS}}{R_\mathrm{QS}}\right)-1 \right|$.The calculation is made assuming a transition of a hadronic type star with a combined EoS model including additional hyperons \citep{bombaci} to a hybrid star with a quark content described by the MIT bag model. Two relativistic non-linear Walecka-type  EoS models of Glendenning-Moszkowsky \citep{glen} are tested, namely, GM1 (blue curve) and GM3 (red curve) usually used to describe the hadronic phase. 
The typical compressibility values $K$ 
for the GM1 model are stiffer than for the GM3 model. Note that in this scenario of hybrid stars the maximum mass for the hybrid object must be somewhat lower than in the case of the pure hadronic star. Although both of them predict maximum hadronic stars masses somewhat on the  low side with $M_{\rm max} \le 1.9 M_{\odot}$, (as compared to recent measurements reporting masses of about $2M_\odot$ \citep{demorest}), they can be considered illustrative to show a general trend of the typical original configurations before transitioning to hybrid or pure QSs. In this figure the value of some nuclear parameters such as  the s-quark mass has been taken to be $m_s=150\, \rm MeV/c^2$ and the strong coupling constant $\alpha_S=0$. Surface tension is $\sigma=30 \, \rm MeVfm^{-2}$. No curvature energy has been considered.

\begin{figure}
\begin{center}
\includegraphics [angle=-90,scale=0.75] {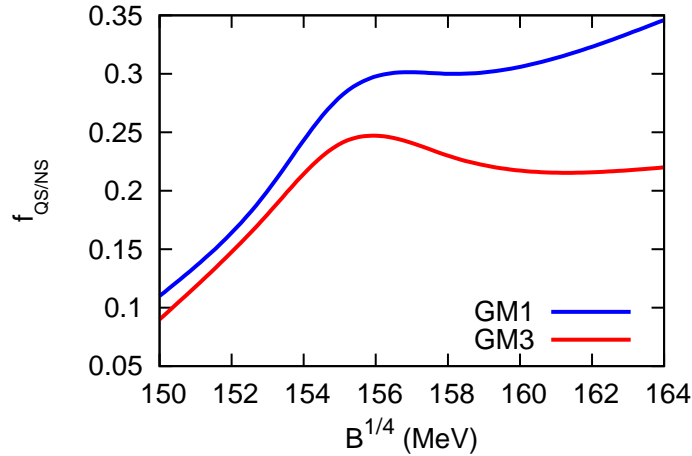}
\caption{Efficiency of the energy extraction by the NS$\to$QS conversion as a function of the bag constant in the combined model of GM1 (blue curve) and GM3 (red curve) and the MIT bag model. See text for details. 
}
\label{fig:eos}
\end{center}
\end{figure}

In this situation the initial configuration masses and radii are not the same, since they depend on the value of the $B$ bag constant  in the MIT model. Following \citet{bombaci} we consider that the transition takes place between a configuration in the initial hadronic star with gravitational mass $M^G_i=\int_0^{R_{NS}} \rho_B(r) d^3{\bf r}$ and with some baryonic mass $M^B_i=\int_0^{R_{NS}} n_B(r) d^3{\bf r}$ and that of a final hybrid object (computed with the hybrid EoS) with the same baryonic mass, $M^B_F=M^B_i$, since that charge is conserved. $\rho_B(r)$ and $n_B(r)$ are the energy density and baryonic mass density for the baryonic object.
Physically, the transition occurs due to the fact that as the density or central pressure exceeds a threshold value, it is energetically allowed that, at the transition point, bubbles of quark matter form. This is due to the lowering of the chemical potential in the system as more degrees of freedom are available. These lumps of quark matter can grow very rapidly, driving a burning front and the original hadronic star will be converted into a hybrid star or quark star. In the mechanism involving self-annihilating DM this energy released could be considered as {\it sparks} that would coalescence or ignite the medium and allow quark deconfinement out of the baryons present (this is B-dependent and therefore, EoS dependent in our model).
If we further include corrections due to slightly different masses for the initial and final configurations we have $f_\mathrm{QS/NS} \approx \left| \Delta E_\mathrm{grav,NS\to QS}/E_\mathrm{grav,prog\to NS} \right| \approx \left|\left(\frac{M_\mathrm{QS}}{M_\mathrm{NS}}\right)^2\left(\frac{R_\mathrm{NS}}{R_\mathrm{QS}}\right)-1 \right|$. We typically find $f_\mathrm{QS/NS}\approx 0.1-0.3$, which illustrates that the transition mechanism discussed here indeed belongs to the class of the most energetic phenomena in the Universe.
The efficiency $f_\mathrm{QS/NS}$ is plotted in Fig~\ref{fig:eos} as a function of the bag constant $B^{1/4}$. It shows that the efficiency $f_\mathrm{QS/NS}$ is higher as the bag constant $B$ grows, pointing to heavier DM particles in order to ignite the nucleation of quark matter lumps but well in the conservative range shown in Fig. 2  in \citet{perez:10}. Further work is needed to clarify the role of the EoS in the efficiency of the transition since the possible existence of a burning front able to fully convert the NS into a QS and proceed towards higher radii is a crucial point.

Numerically, using typical values $M_\mathrm{NS}=1.5\,\mathrm{M_\odot}$, $R_\mathrm{NS}=12\, \mathrm{km}$ and $R_\mathrm{QS}\simeq 7\, \mathrm{km}$ results 
in $\Delta E_\mathrm{grav}\simeq 3.5\times 10^{53}\, \mathrm{erg}$. 
In the NS$\to$QS transition most of the energy $\Delta E_\mathrm{grav}$ will be radiated as neutrinos and photons, as showed by detailed calculations  \citep{jai02,vogt}. Additionally gravitational radiation should also be emitted (see \S\ref{sect:gw}).
We assume now that a small fraction, $f_\mathrm{ej}$, of this energy may be injected into the outer crust which is then ejected and  becomes relativistic. 
The kinetic energy of the expelled crust is
$
E_\mathrm{kin}\simeq f_\mathrm{ej} \Delta E_\mathrm{grav} 
$
if acceleration is complete.
The isotropic equivalent energy  that could be radiated as gamma-rays by such an ejecta can be estimated by 

\begin{equation}
E_\mathrm{\gamma,iso}\simeq 3.5\times 10^{51}\, \left(\frac{f_\mathrm{b}}{100}\right) \left(\frac{f_\mathrm{\gamma}}{0.1}\right)\left(\frac{f_\mathrm{ej}}{10^{-3}}\right)\left(\frac{R_{NS}}{12 \, \rm km}\right)^{-1}\left(\frac{M}{1.5\, M_{\odot}}\right)^2 \, \mathrm{erg}
\, ,
\label{eq:egammiso}
\end{equation}
where $f_\mathrm{\gamma}$ is the efficiency of gamma-ray energy extraction from the ejecta and could range from $\sim 0.01-0.1$ for the extraction of kinetic energy by internal shocks \citep{rees:94,kobayashi:97,daigne:98} to $\sim 0.5$ for photospheric emission \citep[see e.g][]{rees:05} or magnetic reconnection \citep{thompson:94,zhang:11}. 
As discussed below, relativistic motion of the ejecta is favored by a small ejected mass, $M_\mathrm{ej}\sim10^{-5}\, \mathrm{M_\odot}$, which corresponds typically to the outer crust. The crust is defined as the region where  the density drops below nuclear saturation density (NSD) at $\sim 2\times 10^{14}$ $\rm g\,cm^{-3}$, and the outer crust is limited by the neutron drip (ND) at $\sim 4\times 10^{11}\, \mathrm{g\,cm^{-3}}$
 (see \S\ref{sect:engine}).
Then, the value $f_\mathrm{ej}\simeq 10^{-3}$ used in Eq.~\ref{eq:egammiso} would correspond for instance to a situation where  1\% of $\Delta E_\mathrm{grav}$ is injected in the crust and 10\% of this energy is injected preferentially in the outer crust, so that $f_\mathrm{ej}\sim 0.01 \times 0.1$. As there is large uncertainty in this estimate we use a safe approximation below the $f_\mathrm{ej}\sim 0.1$ quoted in other detailed calculations \citep{ouyed05}.

This estimate of $E_\mathrm{\gamma,iso}$ is in reasonable agreement with observations of SGRBs: 
the NS$\to$QS conversion scenario investigated here can reproduce measured energies 
in SGRBs  for $f_\mathrm{b} f_\mathrm{\gamma} f_\mathrm{ej} \simeq  3\times 10^{-4}-0.3$. \\

Regarding the relativistic nature of the ejecta, the maximum Lorentz factor, $\Gamma_\mathrm{max}$, that can be reached depends on the expelled fraction or ejected mass, $M_\mathrm{ej}$, of the crust. The maximum Lorentz factor can be deduced from the estimate of $E_\mathrm{kin}$ above, and from the ejected mass $M_\mathrm{ej}$ as:
\begin{equation}
\Gamma_\mathrm{max}=\frac{E_\mathrm{kin}}{M_\mathrm{ej} c^2} 
\simeq 19\, \left(\frac{f_\mathrm{ej}}{10^{-3}}\right)\left(\frac{R_{NS}}{12 \, \rm km}\right)^{-1}\left(\frac{M}{1.5\, M_{\odot}}\right)^2 \left(\frac{M_\mathrm{ej}}{10^{-5}\, M_\odot}\right)^{-1} \, .
\end{equation}
Again, it seems that Lorentz factors above $15$, in agreement with the observational constraints described above, can be reached as long as the ejected mass remains low ($M_\mathrm{ej} \la 10^{-4} M_\odot$) and the fraction of the energy injected in the outer crust is not too small ($f_\mathrm{ej}\ga 10^{-3}$). Even ultra-high relativistic ejecta with $\Gamma>100$ could in principle be produced, if the ejected mass is really small. This could be the scenario where the {\it outer} crust is expelled.\\

We conclude that the DM self-annihilation triggered NS$\to$QS scenario can, in principle, release enough energy to power a SGRB, and that the ejected crust can reach high Lorentz factors, which is necessary to emit gamma-rays. This is however strongly dependent on the two parameters $f_\mathrm{ej}$ and $M_\mathrm{ej}$ which are very difficult to predict accurately at this stage.
This will be briefly discussed in \S\ref{sect:engine}.\\

\section{Event rate and delay between the NS formation and the transition to a QS}
\label{sect:rate}

Assuming that the transition of a NS to a QS triggered by the self annihilation of accreted DM in the core can inject enough energy into relativistic ejecta to produce a SGRB, it is worth  comparing  the observed rate of these phenomena with the predicted rate of NS$\to$QS conversions.
The local rate per unit volume of SGRBs, $\mathcal{R}_\mathrm{SGRB}$, can be estimated from their observed rate and distribution of flux using a population model (luminosity function, comoving rate, etc), as has been done by \citet{guetta:06}, \citet{nakar:06}, who obtain
\begin{equation}
\mathcal{R}_\mathrm{SGRB} \simeq \left( 400 \to 1500\right) \left(\frac{\left\langle f_\mathrm{b}\right\rangle}{50}\right)\, \mathrm{Gpc^{-3} \,yr^{-1}}\, .
\label{eq:rateSGRB}
\end{equation}
The lower limit corresponds to a comoving SGRB rate that follows the cosmic star formation rate with a long delay, and the upper limit to a constant comoving rate. Unfortunately, the unknown distribution of the beaming factor and its average $\langle f_\mathrm{b}\rangle$ is again a major source of uncertainty in this estimate.\\

In the scenario presented is this work,
the local rate of NS formation, estimated as the local rate of type II supernovae, gives an {\it upper limit} for the rate of NS $\to$ QS conversions. It is of the order of \citep{dahlen:04}
\begin{equation}
\mathcal{R}_\mathrm{NS\to QS,max}^\mathrm{(SNII)} \simeq 5\times 10^5\, \mathrm{Gpc^{-3} yr^{-1}}\, .
\end{equation}
Then the ratio of the former two rates is
\begin{equation}
{\mathcal{R}_\mathrm{SGRB}}/{\mathcal{R}_\mathrm{NS\to QS,max}^\mathrm{(SNII)}} \simeq \left( 8\times 10^{-4} \to 3\times 10^{-3}\right) \left(\frac{\left\langle f_\mathrm{b}\right\rangle}{50}\right)\, .
\label{eq:rateSNII}
\end{equation}
From these estimates only, one can conclude that NS $\to$ QS transition can be much more frequent than SGRBs, depending on the fraction of NS that will experience such a transition. If all SGRBs are due to NS $\to$ QS conversions, the low ratio obtained in Eq.~\ref{eq:rateSNII} could be (i) either due to the fact that only a fraction of NS $\to$ QS transitions lead to short GRBs; (ii) or be related to the fact that only a small fraction of NS are converted into QS.

Having only a small fraction of NS converting to a QS is expected if the delay $\Delta t$ between the formation of a NS and its conversion to a QS is usually long, e.g. $\Delta t\sim$ several Gyr. 
For a broad probability distribution $p(\Delta t)$ of the delay $\Delta t$, with a high mean value $\left\langle\Delta t\right\rangle$ of the order of the Hubble time, most conversions would not have occured yet, resulting in a low ratio  ${\mathcal{R}_\mathrm{SGRB}}/{\mathcal{R}_\mathrm{NS\to QS,max}}$, the observed short GRBs being produced by the conversions with the shortest delays. In addition, such a distribution of $p(\Delta t)$  would make the scenario in good agreement with the properties of short GRB afterglows and host galaxies: contrary to the case of LGRBs which are always observed in central regions of star forming galaxies \citep{bloom:2002}, SGRBs can occur at any place (sometimes at the periphery or outside) in any type of galaxy \citep{berger:11}. 
This indicates a delay between the end of the life of the massive progenitor star and the production of the short GRB.
It is already a well known fact that most NSs are born with a natal velocity kick $v$ that can be as large at $\approx 10^3$ km/s \citep{arzou}. 
Therefore, with high values of $\Delta t$, not only any correlation with star formation is lost, but the NS can  travel a distance 
\begin{equation}
D\approx v \Delta t\approx 10\, \mathrm{kpc} \left(\frac{v}{1000\, \mathrm{km\,s^{-1}}}\right)\left(\frac{\Delta t}{10\, \mathrm{Myr}}\right)\, ,
\label{eq:distance}
\end{equation}
and most likely experience the conversion to a QS far from the
 galactic central regions, or even outside the galaxy. 

In the scenario presented here a lower limit on $\Delta t$ can be obtained from the physics of the transition, which is triggered by self-annihilation of accreted DM from the galactic halo. 
In our Galaxy the DM density distribution can be taken to be of a type \citep{profile}
\begin{equation}
\rho_{\chi}(r)=\rho_{-2}e^{\frac{-2}{\alpha} [ (\frac{r}{r_{-2}})^{\alpha} -1]} 
\end{equation}
with parameters $\rho_{-2}=0.22 \,\rm GeV/cm^3$, $\alpha=0.19$, and $r_{-2}=16 \,\rm kpc$ so that at the solar neighbourhood the Keplerian velocity is $v \sim 220$ km/s and the local DM density is $\rho_{\chi, 0} \sim 0.3\, \rm GeV/cm^3$. 
DM from this halo can be accreted by gravitational capture \citep{Gould} at a rate at peak of NS distribution \citep{lorimer} of
\begin{equation}
 {\cal C} \approx \frac{2.7 \times 10^{29}}{ m_{\chi} (\rm GeV)} \frac{\rho_{\chi}}{\rho_{\chi, 0}}\, \mathrm{particles\, s^{-1}}\, ,
\end{equation}
but off-peak, at about solar circle it may be reduced somewhat. In this estimate, we assume a WIMP-nucleon (spin independent) cross section $\sigma=\sigma_{\chi N}=7\,\times\,10^{-41}\,\rm cm^2$ \citep{bertone}. Once the steady state, resulting from competing processes of annihilation and accretion,  has been reached, the elapsed time is
  $\tau_\mathrm{DM} \approx \frac{1}{\sqrt{ {\cal C}<\sigma_a v>/V}}$ where $V$ is the volume of the star. Typically, if we assume a light  $\sim 20$ $\rm GeV/c^2$ DM particle, as direct detections experiments seem to preliminary suggest, we obtain $\tau_\mathrm{DM}\sim 3.5 \,\times\, 10^3$ yr. For a velocity (temperature) dependent cross section and heavier DM particles $\sim 100$ $\rm GeV/c^2$, this delay can be longer, $\tau_\mathrm{DM} \ge 6 \,\times\,10^5$ yr \citep{kou08}.
From the condition that the transition should likely occur after reaching the steady  state, we get a lower limit on the delay $\Delta t$, 
\begin{equation}
\Delta t_\mathrm{min}\simeq \tau_\mathrm{DM}\simeq 10^3-10^5\, \mathrm{yr}\, .
\end{equation}
Notice that this delay is already much too large to allow to consider producing LGRBs with NS $\to$ QS conversions, as an associated SN is sometimes found in association with such events within a few days \citep{grb-SN}. On the other hand, it is too small to have SGRBs uncorrelated with star formation, which is required by the observed diversity of SGRB host galaxies and afterglow locations \citep{berger:11}.

We conclude that $\Delta t$ is related to the microphysical processes happening in the burning of the NS. This, in turn, depends on the ability of the compact object  (mainly related to the EoS) and environment conditions to accrete DM and, therefore,  cannot be simply predicted and should present some spread and extend up to a large value $\Delta t_\mathrm{max}$ that can be constrained by observations.

In our Galaxy, there are 
hundreds of confirmed NSs, with measured radii and masses \citep{latt}.
On the other hand, the subsample of confirmed NSs with an estimate of the age is very small, due to the difficulty of the age determination,
based on cooling theory \citep{page}.
It seems that NSs with an age of at least $\sim 1-10$ Myr are identified \citep{yako}
which would put a lower limit on the maximum value of the delay $\Delta t$, 
\begin{equation}
\Delta t_\mathrm{max} > 10\, \mathrm{Myr}\, .
\end{equation}
Such delays, as shown by Eq.~\ref{eq:distance}, lead to travelled distance of the order of 10 kpc or more, reaching the outskirts of a galaxy but are still too small to de-correlate from the star formation activity. One can not of course exclude that much older NSs are present in the sample of confirmed NSs, which would increase $\Delta t_\mathrm{max} $ up to 100 My or more. Nevertheless, a detailed comparison with the properties of SGRB host would require the knowledge of the distribution $p(\Delta t)$, which seems yet out of reach. 

Another possible approach to constrain the rate of NS$\to$QS transitions in the scenario proposed here is to focus on the kinematics of observed pulsars.
The pulsar tangential velocity data can be fitted with a bimodal distribution peaked around $v_1=300$ km/s and an upper $v_2=700$ km/s \citep{arzou}. 
The high velocity pulsars with $v>v_2$ roughly 
account for the 10$\%$  of the pulsar population. The first peak in the distribution is believed to be due to the kick velocity given to the NS when it is formed in a core collapse supernova.
It has been suggested by \citet{popov} that the second higher velocity peak could be due to a  {\it second} kick when the NS$\to$QS conversion takes place.
In previous work \citep{perez:10,perez:11}, it was shown that  DM seeding in NSs may form  a stable and long-lived strange quark matter (SQM) lump that could induce a  partial NS conversion into a hybrid SQM star. This event may produce, as a consequence, an effect  on the kinematics with a birth kick and rotation as a result of the partial burning of the NS \citep{perez:11} able to produce up to $v \approx 10^3$ km/s and relative changes in the angular velocity of $\Delta \Omega/\Omega\approx 10^{-3}-10^{-2}$  as a result of the off-center mechanism driving the transition. Other mechanisms quoted in the literature rely on some kind of asymetry \citep{lugones} in the front progress.  
 From this interpretation of the observed pulsar velocity distribution, $\sim 10\%$ could be taken as an upper limit for the frequency of the NS$\to$QS transitions, leading to an upper limit for the NS$\to$QS rate which is more constraining than in Eq.~\ref{eq:rateSNII}, i.e.
%
\begin{equation}
\mathcal{R}_\mathrm{NS\to QS,max}^\mathrm{(kin)} \simeq 5\times 10^4\, \mathrm{Gpc^{-3} yr^{-1}}\, .
\end{equation}
Since this rate remains higher than the SGRB rate, at this stage the only possibility in this scenario is to 
to assume that only a fraction $f_\mathrm{SGRB}\sim 1-10\%$ of NS $\to$ QS transitions produce a short GRB, leading to 
\begin{equation}
\frac{\mathcal{R}_\mathrm{SGRB}}{\mathcal{R}_\mathrm{NS\to QS,max}^\mathrm{(kin)}} \simeq \left( 8\times 10^{-2} \to 3\times 10^{-1}\right) \left(\frac{\left\langle f_\mathrm{b}\right\rangle}{50}\right)\, \left(\frac{\left\langle f_\mathrm{SGRB}\right\rangle}{0.1}\right)\, .
\end{equation}
This additional factor $f_\mathrm{SGRB}$ can be related to internal processes, such as the capacity of the burning front to proceed to the crust (see \S\ref{sect:engine}). 

Note that the value of the efficiency $f_\mathrm{SGRB}\sim 1-10\, \%$ has been obtained here assuming that all short GRBs are produced by NS $\to$ QS transitions. However, we will show below (\S\ref{sect:duration}) that this scenario naturally leads to short GRB durations $T_\mathrm{90}<0.1$ s, so that only very short GRBs are good candidates for counterparts of these NS $\to$ QS transitions. Then, it reduces in principle even more the value of $f_\mathrm{SGRB}$.

On the other hand, observations of afterglows of ultra-short GRBs, identifications of their host galaxy and  measurements of their distance are extremely rare so that their intrinsic rate is unknown. The distribution of the duration of BATSE bursts show that $\sim 8\,\%$ of the short bursts have a duration below 100 ms \citep{horvath:02}. This factor cannot however be directly applied to the estimate of the intrinsic rate given by Eq.~\ref{eq:rateSGRB}. Indeed, this population of very short bursts may very well be a separate group with different properties: they are usually made of a single short and hard spike with possible substructure on a timescale of a few 10 $\mu$s \citep{cline:99}. The analysis ($\left\langle V/V_\mathrm{max}\right\rangle$ and $\log{N}-\log{S}$ diagram) of the distribution of very short bursts observed by BATSE and Konus gives some evidence for a local origin \citep{cline:99,cline:05}. This is however a bit contradictory with more recent results obtained by \textit{Swift}: 10 very short bursts with a duration of less 100 ms have been detected by Swift until June 2012\footnote{Source: \anchor{http://swift.gsfc.nasa.gov/docs/swift/archive/grb_table.html/}{Swift GRB table at \texttt{
http:$\backslash\backslash$swift.gsfc.nasa.gov$\backslash$docs$\backslash$swift$\backslash$archive$\backslash$grb\_table.html
}}}. All of these show indeed a single short duration hard spike in the BAT. In many cases, the afterglow has not been identified or is very weak at the limit of the detection (GRB 050925, GRB 051105A, GRB 070209, GRB 070810B, GRB 070923, GRB 090417A, GRB 100628A, GRB 120305A). There are however two very short bursts where the afterglow has been well detected and localized and where there is a good candidate for the host galaxy. In both cases, the host candidate is an early type galaxy and the afterglow shows a large offset of a few 10 kpc: GRB 050509B seems to be associated to an elliptical galaxy at $z=0.225$ with an offset of 35-55 kpc \citep{SGRB-afterglow,berger:11}; GRB 090515 is probably associated to an early type galaxy at $z=0.403$ with a large offset of 75 kpc \citep{berger:11}. These two examples indicate a population at cosmological distance which is uncorrelated to star formation, in agreement with the discussion above. The sample is however much too small to allow for a determination of the intrinsic rate of very short GRBs and an estimate of $f_\mathrm{SGRB}$.


\section{Central engine and crust masses }
\label{sect:engine}

In this section, we model the basis of the central engine mechanism of the internal burning front that will induce a resultant outflow with several relativistically moving emitting  regions. 
This discussion is based on a series of works
 \citep{cheng,dai,ouyed,pac05,xu,fischer}.  Nuclear processes involving quark deconfinement in the hadronic phase may happen if the $m_{\chi} \ga 1-100$ $\rm GeV/c^2$ DM particle candidate self-annihilates liberating  $\sim \rm MeV-GeV$ photons and other light particle pair products. The cor\-responding lump of strange quark matter ($uds$ matter) evolves towards the formation of a fireball  that may cause stress and tension on the base of the inner crust of a NS. 
Since there are only a few preliminar simulations \citep{miller, herzog} developed so far on the full process of energy transport from the inner deconfined regions to outer regions through a burning front there is not much information where this front may stop or whether it fully proceeds to the outer crust. In these simulations the NS burning condition is dynamically analyzed showing the possibility that a hybrid star may form if the conversion front is not able to proceed as {\it burning}. However, a full detailed treatment has not hitherto been performed. \\

In previous work \citep{perez:11}, it was found that the seeding is most likely a non central process, and therefore, geometrically,  the burning front progression may not be a radially symmetric dynamical process (see Fig.~2 in \citet{perez:12}). 
The possibility of a beamed ejection 
of the outer crust arises then from the anisotropy in the progression of the burning front. This has been somewhat explored in the work of \citet{lugones} where they discuss the possibility of preferred ejection of a fireball 
through the polar caps.
In the present work, we discuss the constraints on the ejected mass as a result of the NS$\to$QS conversion. Similar to what may happen in the heavy ion collision events  in large colliders, a fireball may be formed and grow rapidly. Then, a pure nucleonic EoS of matter, would no longer be valid as nucleon quark content may be deconfined. Since the newer EoS is softened, later evolution may lead to the fact that the original hydrostatic structure is not longer energetically possible and re-adjustement of the object  to favor lower radii and masses and then to build up tension in the crust tending to eject it and break it up. 

The crust mainly corresponds to matter where the density is below the NSD. Therefore for lower densities a myriad of nuclei with mass number $A$ populate this phase. Since the particular distribution of nuclei depend on the competition of short-range hadronic interaction and long range Coulomb interaction a set of irregular shapes different from the spherical one can arise, forming what is known as the {\it pasta phase}. As a result, low density matter in the crust of NSs is mostly neutron-rich due to the deleptonization caused by neutrino escape in the first stages of cooling. Its isospin content is closer to that of neutron matter than to regular nuclei where proton and neutron content are balanced $Z\sim N$.
 From the $^{56}$Fe iron content usually assumed 
 in the lower density region in the crust, there is a series of neutron-rich nuclei going through heavier $^{64}$Ni, $^{82}$Ge, $^{120}$Sr and after this the ND transition takes place, signaling the inner crust of the NS. In the inner crust, 
 surrounding these non-uniform pasta structures, a neutron gas is filling the system, as  has been directly simulated in previous work \citep{pasta2,pasta, termos}. 

\begin{figure}[hbtp]
\begin{center}
\includegraphics [angle=-90,scale=0.75] {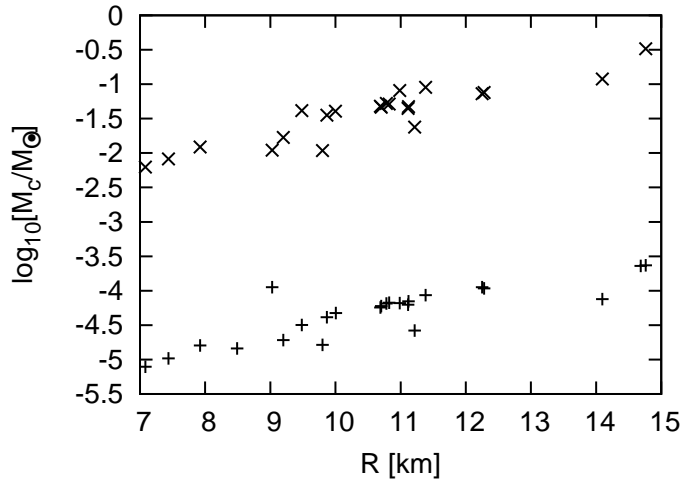}
\caption{Logarithm (in base $10$) of the crust mass (upper points) up to NSD and outer crust  mass (lower points) up to ND for different EoS considered in \citet{Datta} as a function of stellar radius}
\label{Fig3}
\end{center}
\end{figure}

Fragmented emission of ejecta is therefore possible since the stress in the base of the crust from a burning front may liberate the most abundant structures and the rest of lower $A$ nuclei. The gradient of composition (heavy $\to$ light) should also produce  some variability during the crust ejection, all regions of the crust being not necessarily ejected with the same Lorentz factor.  
An additional source of uncertainty is related to the fraction of the initial energy release associated to the transition in the core which will be injected into gamma-ray photons, and the associated photo-disintegration of heavy nuclei. An efficient photo-disintegration would lead to a modified chemical composition biased towards light elements, and would enrich the medium by free neutrons that can be an additional source of energy (neutron decay) and internal dissipation in the ejecta (see \S\ref{sect:duration}).
In order to size the importance of this ejection, 
we 
plot in Fig.~\ref{Fig3} 
the crust mass $M_\mathrm{c}$ 
as a function of the stellar radius. These data are obtained integrating the TOV \citep{tov} equations for eight representative different EoS \citep{Datta} up to the core. We compute the mass of the crust by integrating above a critical density, either the neutron drip density $\rho_\mathrm{ND} \approx 4\,10^{11}$ $ \rm g/cm^3$ to estimate the mass of the outer (solid) crust only, or the nuclear saturation density 
$\rho_\mathrm{NSD}\sim 2\times 10^{14}$ $\rm g/cm^3$ 
(where crust-core boundary sets in)  to estimate the mass of the entire crust.
We find that the condition $M_\mathrm{ej}\la 10^{-4}\, M_\odot$  needed for the relativistic motion (\S\ref{sect:energetics}) can be 
fulfilled if only the outer crust is ejected ($M_\mathrm{c}\sim 10^{-5}\mathrm{M_\odot}-10^{-3.5}\, \mathrm{M_\odot}$), whereas it becomes much more difficult if the whole crust is expelled ($M_\mathrm{c}\sim 10^{-2.5}\mathrm{M_\odot}-10^{-0.5}\, \mathrm{M_\odot}$).
The details of the physics of the propagation of the burning front and the associated energy deposition are required to estimate precisely the fraction of the crust which is ejected. If this fraction varies from a NS$\to$QS transition to another, this would naturally lead to $f_\mathrm{SGRB}<1$ as discussed in the previous section.  
\\


\section{GRB Duration, Light curve, Spectrum}
\label{sect:grb}
In this section we discuss some expected properties for 
SGRBs produced in the scenario proposed in this work. We focus on the identification of distinct features related to the specific central engine that could help in distinguishing this mechanism from others proposed in the literature.

\subsection{Duration and light curve}
\label{sect:duration}

Let us assume that the ejected outer crust has a mass 
 $M_\mathrm{ej}=M_\mathrm{ej,-5}  \times 10^{-5}\, M_\odot$, an initial energy $E_\mathrm{ej}=f_\mathrm{ej,-3}\times 3.5\times 10^{50}\, \mathrm{erg}$ and a width $\Delta = c \Delta t$, where $\Delta t=\Delta t_{-6} \times 10^{-6}\, \mathrm{s}$ is the light crossing time ($\Delta =\Delta t_{-6}\times 300\, \mathrm{m}$). The maximum final Lorentz factor that can be reached in the ejection is
\begin{equation}
\Gamma
 = \frac{E_\mathrm{ej}}{M_\mathrm{ej}c^2} = 20\, M_\mathrm{ej,-5} ^{-1} f_\mathrm{ej,-3}\,.
\end{equation}
Due to the large physical uncertainties regarding the energy deposition in the crust and the following ejection, we do not attempt here a detailed calculation of the relativistic ejection \citep[see e.g.][for a self-similar solution of the propagation of a strong shock wave within the NS and the following shock breakout]{pan}.
For a thermal acceleration, the saturation to this value of the Lorentz factor will occur at radius
 \begin{equation}
R_\mathrm{sat} \simeq
\Gamma 
R \simeq 2
\times 10^7\, M_\mathrm{ej,-5} ^{-1} f_\mathrm{ej,-3}\, \mathrm{cm}\, ,
\end{equation}
assuming an initial radius $R_\mathrm{ej}=R-\Delta\simeq R_\mathrm{NS}=12$ km for the ejection. The ejecta will become transparent for its own radiation at the photospheric radius
\begin{equation}
R_\mathrm{ph} \simeq \sqrt{\frac{\kappa M_\mathrm{ej}}{4\pi}} \simeq 2\times 10^{13}\, M_\mathrm{ej,-5}^{1/2}\, \mathrm{cm}\, ,
\end{equation}
where we take the Thomson opacity $\kappa\simeq
\, 0.2 \,\rm cm^2/g$. Here we assume $Y_\mathrm{e}=0.5$ free electron per nucleon in the expanding gas, whereas the dynamical chemical composition discussed in the previous section may lead to lower values. This does not affect too much the discussion as the dependence goes moderately as $R_\mathrm{ph}\propto \kappa^{0.5}$. 

This expression of $R_\mathrm{ph}$  is valid for 
$R_\mathrm{ph}\gg R_\mathrm{is} \simeq 2\Gamma^2\Delta$. The radius $R_\mathrm{is}$ is defined below and is of the order of a few $10^7$ cm so that the condition is always true for the typical parameters considered here.
The 
estimates of $R_\mathrm{sat}$ and $R_\mathrm{ph}$ are based on standard fireball theory for GRBs \citep[see e.g.][]{piran:04} and are only rough estimates for the scenario considered here as the physics of the ejection of the crust is much more complex. 
We should
 bear in mind that the ejecta composition \citep{kumiko} and fragmentation may affect both the dynamics ($R_\mathrm{sat}$) and the interaction with radiation ($R_\mathrm{ph}$) \citep {ouyed}. Therefore, if the energy is not deposited in an homogeneous way in the expelled crust, the final Lorentz factor in the ejecta may not be uniform. In addition, fragmentation of the crust during its ejection will also lead to some variability in the ejecta. If variations of $\Gamma$ are present on length scales $c t_\mathrm{var} < \Delta$, this will induce collisions (internal shocks) that will dissipate energy at a typical radius $R_\mathrm{is}$ given by
\begin{equation}
R_\mathrm{is} \la 2 \Gamma^2 \Delta \simeq 2\times 10^7\, M_\mathrm{ej,-5} ^{-2} f_\mathrm{ej,-3}^2 \Delta t_{-6}\,  \mathrm{cm}\, .
\end{equation} 

Clearly,  this possible internal dissipation will always occur well below the photosphere, and, depending on the value of the Lorentz factor, even before the acceleration is complete. Most of the dissipated energy should contribute again to the acceleration and the internal shock phase will only tend to smooth out the initial internal variability, without contributing to the emission.

\begin{figure}[hbtp]
\begin{center}
\includegraphics [scale=0.5] {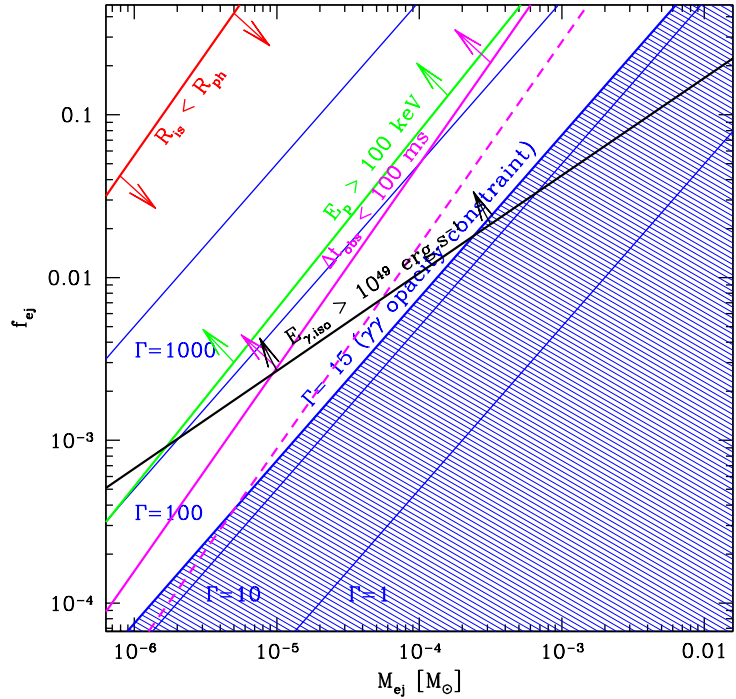}
\caption{In the parameter space mass of the ejected outer crust $M_\mathrm{ej}$ vs efficiency of the energy injection in the crust $f_\mathrm{ej}$, the following constraints are plotted: (1) lines of constant Lorentz factor are plotted in blue for $\Gamma=1$ (non relativistic limit), $10$, $100$ and $1000$. The limit $\Gamma\simeq 15$ obtained from the compactness argument (see \S\ref{sect:energetics} is plotted in thick line and the forbidden region is shaded; (2) the limit where the radius of the internal dissipation $R_\mathrm{is}$ is equal to the radius of the photosphere is plotted in red (most of the parameter space is well below this line, i.e. $R_\mathrm{is}\ll R_\mathrm{ph}$); the limit where the observed duration of the prompt emission of photospheric origin is equal to 100 ms (1 s) is plotted in magenta solid (dashed) line; the limit where the observed peak energy of the photospheric emission is 100 keV is plotted in green; the limit where the isotropic equivalent gamma-ray energy emitted by the photosphere is equal to $10^{49}\, \mathrm{erg.s^{-1}}$ is plotted in black. The effect of the redshift of the source on the duration and the spectrum are not included. Other parameters are $\Delta t_{-6}=1$ (i.e. the width of the outer crust is 300 m) and $f_\mathrm{b}=50$ (i.e. the ejecta is beamed within $\sim 10^\circ$). As can be observed, a large fraction of the parameter space (top-left region) corresponds to ultra-relativistic outflows ($\Gamma\ga 100$) producing a very short ($<100$ ms) but bright ($E_\mathrm{\gamma,iso}>10^{49}\, \mathrm{erg}$) spike of hard ($E_\mathrm{p}>100$ keV) photons, i.e. a very short GRB.}
\label{fig:grbnsqs}
\end{center}
\end{figure}

The ejecta will expand freely initially, but will eventually be decelerated by the external medium. However, due to the expected delay between the formation of the NS and the NS $\to$ QS transition, this external medium can correspond to the periphery of the host galaxy or even the surrounding intergalactic medium, i.e. have a low density. We assume a uniform medium with a number density $n=n_{-3}\times 10^{-3}\, \mathrm{cm^{-3}}$. Then the deceleration will start at radius
\begin{equation}
R_\mathrm{dec} \simeq \left(\frac{3 M_\mathrm{ej}^2 c^2}{4\pi E_\mathrm{ej} n m_\mathrm{p}}\right)^{1/3} 
\simeq 
4\times 10^{17}\,
M_\mathrm{ej,-5}^{2/3}
f_\mathrm{ej,-3}^{-1/3} n_{-3}^{-1/3}\,
  \mathrm{cm} \, .
\end{equation}
$m_p$ is the proton mass. From these different estimates, we find for the proposed scenario that
\begin{equation}
R_\mathrm{sat}\la R_\mathrm{is} \ll R_\mathrm{ph}\ll R_\mathrm{dec}\, .
\end{equation}
This would naturally lead to two episodes of emission, a single spike emitted at the photosphere followed by an afterglow starting at late time due to the high value of $R_\mathrm{dec}$. The duration of the prompt spike should be fixed by the intrinsic curvature of the emitting region and its lateral extension, which gives
\begin{equation}
\Delta t_\mathrm{obs} \simeq \min{\left( \frac{R_\mathrm{ph}}{2\Gamma^2 c};\, \frac{\theta_\mathrm{j}^2 R_\mathrm{ph}}{2 c}\right)}
\simeq \min{\left(M_\mathrm{ej,-5}^2  f_\mathrm{ej,-3}^{-2} ;  \left(\frac{\theta_\mathrm{j}}{3^\circ}\right)^2\right)}\, \times\, 0.8\, M_\mathrm{ej,-5}^{1/2}\, \mathrm{s}\, .
\end{equation} 
Except if the ejection is highly beamed, the minimum is usually given by the first term. Nevertheless, this estimate clearly points out towards short ($<1$ s) and even probably very short ($< 100$ ms, see Figure~\ref{fig:grbnsqs})
GRBs without any strong variability (one main single spike).
Due to the low external density, the afterglow should rise slowly and reach a maximum around $t_\mathrm{dec}=R_\mathrm{dec}/2\Gamma^2 c \simeq 2\,\times\,10^4\, M_\mathrm{ej,-5}^{8/3}
f_\mathrm{ej,-3}^{-7/3} n_{-3}^{-1/3}\,
  \mathrm{s}$, i.e. a few 10 ms to a few hours after the GRB, depending on the external density. In addition, the combination of a low external density and a moderate energy will naturally lead to a weak afterglow. Note that if free neutrons are present in the ejecta, due to efficient photo-disintegration (see \S\ref{sect:engine}), these neutrons will decay at a typical radius $R\simeq \Gamma c \tau_{\beta}\simeq 5.4\times 10^{14}\, M_\mathrm{ej,-5} ^{-1} f_\mathrm{ej,-3}\, \mathrm{cm}< R_\mathrm{dec}$, fixed by the mean lifetime $\tau_{\beta}\simeq 900\, \mathrm{s}$, which could  lead to an early additional signature at  $t_\mathrm{obs}\simeq  20\, M_\mathrm{ej,-5} f_\mathrm{ej,-3}^{-1}\, \mathrm{s}$ \citep[see e.g.][]{beloborodov:03}.

\subsection{Spectrum}
\label{sect:spectrum}
If most of the gamma-rays are produced at the photosphere, one should expect a thermal (quasi-Planckian) spectrum, possibly modified at high energy by comptonization \citep{rees:05,peer:2006,beloborodov:11}. 
The peak energy of the observed spectrum will then be located at $E_\mathrm{p}\simeq 3.9\,
k_\mathrm{B} T_\mathrm{ph}$, where $T_\mathrm{ph}$ is the temperature of the photosphere. It can be computed assuming an adiabatic radial expansion from the ejection to the photosphere:
\begin{equation}
E_\mathrm{p}\simeq 3.92 \left(\frac{3 E}{16\pi a R_\mathrm{NS}^2 \Delta}\right)^{1/4} \left(\frac{R_\mathrm{ph}}{R_\mathrm{sat}}\right)^{-2/3}\simeq 
18\, M_\mathrm{ej,-5}^{-1} f_\mathrm{ej,-3}^{11/12} \Delta t_{-6}^{-1/4}\, \mathrm{keV}\, .
\end{equation}

The efficiency of the photospheric emission can also be deduced from the adiabatic evolution and equals $f_\mathrm{\gamma}\simeq \left(R_\mathrm{ph}/R_\mathrm{sat}\right)^{-2/3}$. This leads to a better estimate of the isotropic equivalent energy radiated in gamma-rays,
\begin{equation}
E_\mathrm{\gamma,iso} \simeq 2\times 10^{48}\, \left(\frac{f_\mathrm{b}}{50}\right) M_\mathrm{ej,-5}^{-1} f_\mathrm{ej,-3}^{5/3}\, \mathrm{erg/s}\, . 
\end{equation}
that stands on the lower values for GRBs. Of course, there is still a large uncertainty due to the unknown factors $f_\mathrm{b}$, $ f_\mathrm{ej}$ and $M_\mathrm{ej}$. However, these two estimates confirm the capacity to produce bright and hard spikes of gamma-rays in the considered scenario, as illustrated in Figure~\ref{fig:grbnsqs} where all the constraints on the prompt gamma-ray emission expected in the NS $\to$ QS conversion scenario are summarized.

Another possible signature of the scenario would be the presence of spectral features and lines associated to the specific chemical composition (heavy elements) of the ejected material. 
Following  the work by \citet{mes98} where $\Gamma \simeq 10-100$ were considered, the spectrum could be influenced below the MeV range.

\subsection{Gravitational waves and other non-photonic signatures}
\label{sect:gw}

In addition to the expected short duration burst of $\gamma$-rays and X-rays and to the afterglow, it is likely that a multi-messenger approach must be followed to spot this progenitor scenario for SGRBs. The emission of gravitational waves by the merger of NS+NS or NS+BH binary is the most promising source for detectors such as Virgo and LIGO and their advanced versions \citep{advancedvirgo,harry}, or the Einstein telescope \citep{hild}. The predicted emission during the three main stages of the merger has been extensively studied in the literature \citep[see][for a review]{hughes:09}.
 In the inspiral stage, the signal is very well known up to the last stable orbit. Then, the emission during the merger is more uncertain and must be modeled using supercomputer simulations
 which provide information about the gravitational waveform \citep
{duez,baiotti:10}
(see the recent simulations by \citet{shibata:06,gw2,baiotti:08,anderson,liu:08,giacomazzo:09,
rezzolla:10,kiuchi:10,giacomazzo:11}
which compute the evolution of the merger up to black hole formation, including or not magnetic fields, realistic EOS, etc.). 
The final ringdown stage is also very well known as the inspiral. Therefore, the detection of gravitational waves in association with a SGRB would undoubtedly prove if the progenitor is a merger or not. With a horizon of $\sim 200$ Mpc for NS+NS mergers and $\sim 420$ Mpc for NS+BH mergers, there may be merger detections with advanced Virgo, LIGO\citep{harry} (0.4 to 400 mergers/yr for NS+NS, a slightly smaller rate for NS+BH, \citealt{gwrates}). However, the simultaneous detection with a SGRB if the merger scenario is correct is more uncertain, both for theoretical (beaming of the gamma-ray emission) and instrumental (localization of SGRBs) reasons. 

  The gravitational wave signature of the scenario studied here is not as well known as for compact binary mergers. It is expected that a change in the moment of inertia is  caused by deformations in the NS$\to$QS transition.  Some  preliminary estimates of the transient gravitational wave signal from an explosive quark-hadron phase transition have been done \citep{staff}. It could in principle be detected out to 20 Mpc with advanced Virgo/LIGO, which makes unfortunately the probability of detection with an associated SGRB quite low, even if the intrinsic rate of very short GRBs is quite uncertain (see \S\ref{sect:rate}). 

The strategy for the detection of gravitational waves associated to a GRB is to 
combine searches,  already having identified typical  patterns \footnote{Source:{http://www.ligo.org/science/GW-Burst.php}}  in temporal and directional coincidence with SGRBs that had sufficient gravitational-wave data available,  although it is a very challenging task \citep{abadie2010}. 

An additional signal that may be used to discriminate between progenitor models for SGRBs is the neutrino emission. The specific signal expected from NS+NS mergers has been studied using  supercomputer simulations \citep[see e.g.][]{dessart:09}. 
In this work and within the present scenario we do not attempt to describe such a neutrino emission since it crucially depends on the central engine details. However, it should be expected a neutrino flux originating from hadronic reactions happening in the central engine (conversion in the nucleon burning front of quark deconfinement and strangeness production) or in the ejecta (photo-meson interactions between shock-accelerated protons/nuclei and gamma-ray photons).

In addition, if produced neutrons do not interact they will decay being accompanied by anti-neutrinos. Emission of prompt muon neutrino fluxes have been performed in some GRB general scenarios \citep{Baerwald} and seem to be testable as recent preliminary estimated sensitivities for Ice Cube 86-strings quote values of $E^2 \Phi (E) \sim 5\,\times\, 10^{-8} {\rm GeV\,cm^{-2}\,s^{-1} \, sr^{-1}}$.
Additionally, experiments as AMANDA-II and IceCube have the capability \citep{ama} of detecting anisotropies from the emission \citep{ani} of neutrinos from gamma ray induced air showers.
Therefore, as motivated, the discovery of high energy neutrinos in correlation with a SGRB and GW signal would help disentangling the current scenario. 

\section{Conclusions}
\label{sect:conclusions}

In this work we discuss the possibility that a NS$\to$QS transition may be a central engine model for short ($<$ 1 s) or more probably very short ($<100$ ms) GRBs. This is an alternative to the popular NS+NS or NS+BH merger scenario for SGRBs. Note that this scenario is known to be compatible with observations of SGRBs but has not been proved yet, and that in addition, 
there are a few observations suggesting that very short bursts could be a separate population. We suggest that these very short bursts could be produced by the ejection and acceleration to relativistic speed of the (outer) crust of the NS during the conversion to a QS. We find that
\begin{itemize}
\item The isotropic equivalent gamma-ray energy  $\sim 10^{48}-10^{52}$ can be accounted for, assuming that $0.1-1\%$ of the gravitational energy released by the transition is injected in the outer crust with, however, a large uncertainty due to the unknown beaming factor of the ejection;
\item High Lorentz factors, necessary for the emission of $\gamma$-rays on short time-scales (compactness problem), can be reached, as long as the mass of the expelled crust is less than $\sim 10^{-4} \, M_\odot$; this is attainable for the outer crust in NS models.

\item The rate of very short GRBs can probably be reproduced, assuming that only a fraction of transitions lead to a GRB and that there is on average a large delay between the formation of the NS and the conversion into a QS. Such a delay is expected due to the fact that the transition is triggered by self-annihilation of DM, which needs first to be accreted in the core of the NS. Such delays would then naturally at least partially suppress the correlation between star formation and very short GRBs, which should be observed in any type of galaxy with a large offset (at the present time, there are only two host galaxies of very short GRBs which have been possibly identified, both being early-type, and the offset is a few 10 kpc in both cases);
\item  The prompt gamma-ray emission should be mainly produced at the photosphere, without a strong variability which should be washed out by internal dissipation at much smaller radii.

\item For a large fraction of the parameter space, 
a hard ($E_\mathrm{p}\ga 100$ keV) and short (duration $<$ 100 ms) spike of gamma-rays is expected, in general agreement with observations of very short GRBs;
\item The afterglow should rise slowly and be rather weak, due to a low density of the external medium. A possible additional signature can be expected at early time if the ejected material initially contains free neutrons.
\item Fragmentation of the ejecta arises naturally in this model since non-uniform nuclei arranged in lattice or even pasta phases are present in the low density matter.
\item Spectral features of heavy nuclei relativistically accelerated are expected below $\sim 1$ MeV. 
\end{itemize}

To summarize, possible {\it signatures} compared to the binary (NS, BH) merger scenario are  the shortness of the prompt gamma-ray emission, with possibly a thermal spectrum and spectral features due to the heavy composition,
the associated GW emission, and possibly the properties of the host galaxies and the distribution of the afterglow position in the host, this latter signature being however difficult to characterize due to the uncertainties on the typical delay between the NS formation and the transition to a QS. Clearly, a multi-wavelength/multi-messenger approach is needed to reach a firm conclusion.

\acknowledgments

We thank the MICINN (Spain) MULTIDARK, FIS-2009-07238, FIS2011-14759-E and FIS2012-30926 projects and the  ESF-funded COMPSTAR project for partial financial support. We acknowledge useful discussions with  I. Bombaci, J. Horvath, K. Kotera,  R. Ouyed and C. Providencia. M. A. P. G. would like to thank the IAP for its kind hospitality.


\end{document}